\documentclass[letterpaper,fleqn,12pt]{article}
\usepackage{fullpage,graphicx,physics,amsfonts,amssymb}
\usepackage[square,sort,comma,numbers,sort&compress]{natbib}
\usepackage[hidelinks,colorlinks=true,citecolor=blue,linkcolor=blue]{hyperref}

\newcommand{\Ra}{{\rm Ra}}
\newcommand{\Sc}{{\rm Sc}}
\newcommand{\eqn}[2]{\begin{gather}
#1
\label{#2}
\end{gather}
}
\title{\bf Anomalous mass diffusion in a binary mixture and Rayleigh-B\'enard instability}
\author{\bf 
A. Barletta$^1$, B. Straughan$^2$\\[3pt]
{\small $^1$Department of Industrial Engineering, Alma Mater Studiorum Universit\`a di Bologna,}\\[-3pt]
{\small Viale Risorgimento 2, 40136 Bologna, Italy}\\
{\small $^2$Department of Mathematics, University of Durham,}\\[-3pt]
{\small Durham, DH1 3LE, U.K.}
}
\date{}

\begin{document}
\maketitle

\begin{abstract}\noindent\footnotesize 
The onset of the Rayleigh-B\'enard instability in a horizontal fluid layer is investigated by assuming the fluid as a binary mixture and the concentration buoyancy as the driving force. The focus of this study is on the anomalous diffusion phenomenology emerging when the mean squared displacement of molecules in the diffusive random walk is not proportional to time, as in the usual Fick's diffusion, but it is proportional to a power of time.
The power-law model of anomalous diffusion identifies subdiffusion when the power-law index is smaller than unity, while it describes superdiffusion when the power-law index is larger than unity. This study reconsiders the stability analysis of the Rayleigh-B\'enard problem by extending the governing equations to include the anomalous diffusion.\\[12pt]
{\bf Keywords}\qquad{}Natural convection; Rayleigh-B\'enard problem; Linear stability analysis;\\ Energy
method; Mass diffusion
\end{abstract}

\section{Introduction}
Anomalous mass diffusion refers to situations where the typical characteristics deviate from classical models based on Fick's law. Scenarios and phenomena associated with anomalous mass diffusion are classified as subdiffusion and superdiffusion. In subdiffusion, solute molecules move more slowly than predicted by classical diffusion. In superdiffusion, particles move more rapidly than predicted by classical diffusion. A typical cause of superdiffusion is the long-range interactions between molecules leading to occasional Lévy flights. In fact, Lévy walk is a type of random walk where step sizes are drawn from a Lévy distribution, whose main feature is heavy tails in comparison with the typical Gaussian distribution. There is a wide literature documenting the phenomenology and mathematical modelling of the anomalous diffusion. We refer the reader to the excellent reviews by  Metzler and Klafter \cite{metzler2000random} and by Henry, Langlands and Straka \cite{henry2010introduction}. 

The physical basis of anomalous diffusion relies on the modelling of the molecular random walk process undergone by the solute molecules in a binary mixture. In fact, the relationship between the mean squared displacement $\sigma_x^2$ of the molecules and time $t$ may be sublinear, linear or superlinear, namely
\eqn{
\sigma_x^2 = 2 D t^r ,
}{1}
where $D$ is the generalised diffusion coefficient (with SI units ${\rm m^2/s}^r$) and $r$ is the anomalous diffusion index. Such an index can be $r < 1$ meaning subdiffusion, $r=1$ leading to the usual linear model of diffusion or $r>1$ meaning superdiffusion. 

There are experimental studies and measurements of anomalous diffusion in a plethora of cases involving, for instance, polymers, biopolymers, proteins and liquid crystals, leading to the evaluation of the generalised diffusion coefficient and the power-law index. Examples are the papers by Banks and Fradin \cite{banks2005anomalous} and by Martin, Jerolmack and Schumer \cite{martin2012physical}. We report that fluorescence correlation spectroscopy technique has been employed by many authors \cite{weiss2004anomalous, vercammen2008measuring} for the experimental analysis of anomalous diffusion.

The classical Rayleigh-B\'enard problem has alternative versions where mass diffusion emerges in combination with heat transfer or with the destabilising contribution of one or more concentration gradients. Early investigations of the solutal Rayleigh-B\'enard problem were published several decades ago \cite{veronis1968effect, baines1969thermohaline}. In recent years, this topic has been revisited by other authors \cite{raghunatha2019effect, saleh2020effect, raghunatha2021triple, ali2022soret, amar2024weakly} pointing out new scenarios such as triple diffusion, viscoelasticity, Soret effect and viscous dissipation. The problem of double diffusion is of immense importance due to many applications, such as to see ice melting \cite{carr2003model}, convective motion in volcanic caldera \cite{DECAMPOS2004227, DECAMPOS2008131}, and to renewable energy generation via solar ponds \cite{dineshkumar2022experimental}.

The aim of this paper is reconsidering the stability analysis of the Rayleigh-B\'enard system, under conditions of either subdiffusion or superdiffusion. The analysis is first approached with the linearised set of governing equations for the perturbations and a normal mode analysis. Then, the energy method is employed to approach nonlinearly the case of superdiffusion. The energy method is also employed to reconsider the linear analysis of the instability for the case of subdiffusion under a broader perspective not based on normal modes. This paper extends previous analyses referring to Darcy's flow in a porous medium \cite{barletta2023rayleigh,barstra2023rayleigh}.

\section{The mathematical model}
We consider a horizontal fluid layer with thickness $H$ and infinite horizontal width, bounded by two horizontal planes $z=0$ and $z=H$. The $z$ axis is vertical while the $x$ and $y$ axes are horizontal. The fluid is a binary isothermal mixture. Following the anomalous diffusion model discussed by Henry, Langlands and Straka \cite{henry2010introduction} and employed in the recent studies by Barletta \cite{barletta2023rayleigh} and by Straughan and Barletta \cite{barstra2023rayleigh},  we introduce an effective time-dependent diffusivity, $D r t^{r-1}$, based on the generalised diffusion coefficient $D$. 

By adopting the Oberbeck-Boussinesq approximation, we can express the local mass, momentum and solutal concentration balance equations as
\eqn{
\pdv{{u}_j}{{x}_j} = 0, \nonumber\\
 \pdv{{u}_i}{t} + {u}_j\; \pdv{u_{i}}{x_j} = -  \beta \left( c - c_0\right) {g}_i - \frac{1}{\rho_0}\; \pdv{{p}}{x_i} +  \nu \, \nabla^2 {u}_i  , \nonumber\\
 \pdv{c}{t} +  u_{j}\; \pdv{c}{x_j} =  D\, r\, t^{r-1} \, \nabla^2 c .
}{2}
Here, Einstein's notation for implicit summation over repeated indexes has been adopted. Furthermore, the fields expressing the velocity, dynamic pressure, and concentration have been denoted as $u_i$, $p$ and $c$, respectively. The dynamic pressure is the difference between the pressure and the hydrostatic pressure, while symbol $c_0$ is the reference concentration employed for the definition of the buoyancy force.
Fluid properties $\beta$, $\rho_0$ and $\nu$ denote the mass expansion coefficient, the reference density and the kinematic viscosity, respectively. The gravitational acceleration is given by $g_i = - g\, \delta_{i3}$, where $g$ is the modulus of $g_i$ and $\delta_{i3}$ is the $i3$ component of Kronecker's delta, namely, the $i$ component of the unit vector along the $z$ axis.

Given that the boundaries $z=0,H$ are assumed with uniform concentrations $c_1$ and $c_2$, respectively, there are different combinations of stress-free and rigid boundary conditions for the velocity that are usually employed. For the sake of simplicity, we will hereafter assume stress-free boundary conditions at both boundaries, as this is the only case leading to an analytical solution in the standard diffusion case, $r=1$.  

\subsection{The basic solution}
A stationary basic solution of \eqref{2} satisfying the boundary conditions can be found with the velocity, dynamic pressure gradient and solute concentration expressed as
\eqn{
u_{bi} = 0 \qc \pdv{p_b}{x_i} = - \rho_0 \left( c_b - c_0 \right) g_i \qc c_b = c_1 - \frac{c_1 - c_2}{H}\, z,
}{3}
where the subscript $b$ stands for basic state. This is exactly the basic state considered in the traditional Rayleigh-B\'enard problem with standard diffusion, as the value of $r$ does not influence the solution.

\subsection{Perturbing the basic state}
We introduce a small-amplitude perturbation of the basic state by defining
\eqn{
u_i = u_{bi} + U_i \qc p = p_b + P \qc c = c_b + C,
}{4}
where $(U_i ,P, C)$ are small perturbation fields.
By substituting equations \eqref{3} and \eqref{4} into \eqref{2} and by neglecting the nonlinear terms in the perturbations, we obtain 
\eqn{
\pdv{{U}_j}{{x}_j} = 0,
\nonumber\\
 \pdv{{U}_i}{t} = -  \beta \, {C} \, {g}_i - \frac{1}{\rho_0}\; \pdv{{P}}{x_i} +  \nu \, \nabla^2 {U}_i  ,
\nonumber\\
 \pdv{C}{t} -  {W}\, \frac{c_1 - c_2}{H}  =  D\, r\, t^{r-1} \, \nabla^2 {C} ,
}{5}
where $W$ denotes the $z$ component of $U_i$, while the $x$ and $y$ components will be denoted as $U$ and $V$. Thus, the boundary conditions can be expressed as
\eqn{
z = 0, H :\qquad W = 0 = C , \quad  \pdv{U}{z} = 0 = \pdv{V}{z}.
}{6}
Equations \eqref{5} and \eqref{6} can be rewritten in a dimensionless form by means of the scaling
\eqn{
\frac{U_i}{H/\tau} \to U_i, \quad \frac{C}{c_1 - c_2} \to C, \quad \frac{P}{\rho_0 \nu/\tau} \to {P} \qc
\frac{x_i}{H} \to x_i, \quad \frac{t}{\tau} \to t,
}{7}
where $\tau = (H^2/D)^{1/r}$ is the diffusion time constant. Thus, equations \eqref{5} yield
\eqn{
\pdv{U_j}{{x}_j} = 0,
\nonumber\\
\left( \frac{1}{\Sc} \, \pdv{}{t} - \nabla^2 \right) U_i = \Ra\, C \, \delta_{i3} - \pdv{P}{x_i}   , 
\nonumber\\ 
\pdv{C}{t} - W  = r\, t^{r-1} \nabla^2 C ,  
}{8}
with the boundary conditions
\eqn{
z = 0, 1 : \qquad W = 0 = C \qc \pdv{U}{z} = 0 = \pdv{V}{z}.
}{9}
In equations \eqref{8}, the Schmidt number and the Rayleigh number are defined as
\eqn{
\Sc = \frac{\nu}{D \, \tau^{r-1}} \qc \Ra = \frac{g \beta (c_1 - c_2) H^3}{\nu D\, \tau^{r-1}} .
}{10}
While $\beta$ and $c_1 - c_2$ can be either positive or negative, the inequality $\Ra>0$ addresses a case where the basic state features a potentially unstable solutal stratification. 
By taking the curl of the second equation \eqref{8}, one can get rid of the dynamic pressure gradient, $\partial P/\partial x_i$, so that \eqref{8} can be reformulated in terms only of the fields $W$ and $C$, namely
\eqn{
\left( \frac{1}{\Sc} \, \pdv{}{t}- \nabla^2 \right) \nabla^2 W  = \Ra \, \hat\nabla^2 C  ,
\nonumber\\ 
\pdv{C}{t} - W  = r\, t^{r-1} \nabla^2 C ,  
}{11}
with boundary conditions
\eqn{
z = 0, 1 :\qquad W = 0 = C \qc \pdv[2]{W}{z} = 0 .
}{12}
In \eqref{11}, the operator $\hat\nabla^2$ stands for the two-dimensional Laplacian,
\eqn{
\hat\nabla^2 = \pdv[2]{}{x} + \pdv[2]{}{y} .
}{13}

\section{Normal mode analysis}\label{antnormod}
Let us test the time evolution of normal mode solutions of \eqref{11} and \eqref{12}, namely 
\eqn{
W = f(t) \, \sin(k x) \, \sin(n \pi z) \qc C = h(t) \, \sin(k x) \, \sin(n \pi z) \qc \quad n = 1, 2, 3, \ \ldots\ ,
}{14}
chosen so that the boundary conditions \eqref{13} are identically satisfied. We stress that \eqref{14} describes wavelike modes propagating in any horizontal direction. In fact, the $x$ axis is chosen arbitrarily as no preferred horizontal direction exists for the system. 
Substitution of \eqref{14} into \eqref{11} yields
\eqn{
\frac{1}{\Sc} \, \dv{f}{t} + a_{n,k}^2 \, f  - \frac{k^2 \Ra}{a_{n,k}^2}\; h = 0  ,
\nonumber\\ 
\dv{h}{t} - f  + r\, t^{r-1} a_{n,k}^2\, h = 0 ,  
}{15}
where the shorthand notation, 
\eqn{
 a_{n,k}^2 = n^2 \pi^2 + k^2 ,
}{16}
is employed.

\subsection{The limit of infinite Schmidt number}
\begin{figure}[t]
\centering
\includegraphics[width=0.5\textwidth]{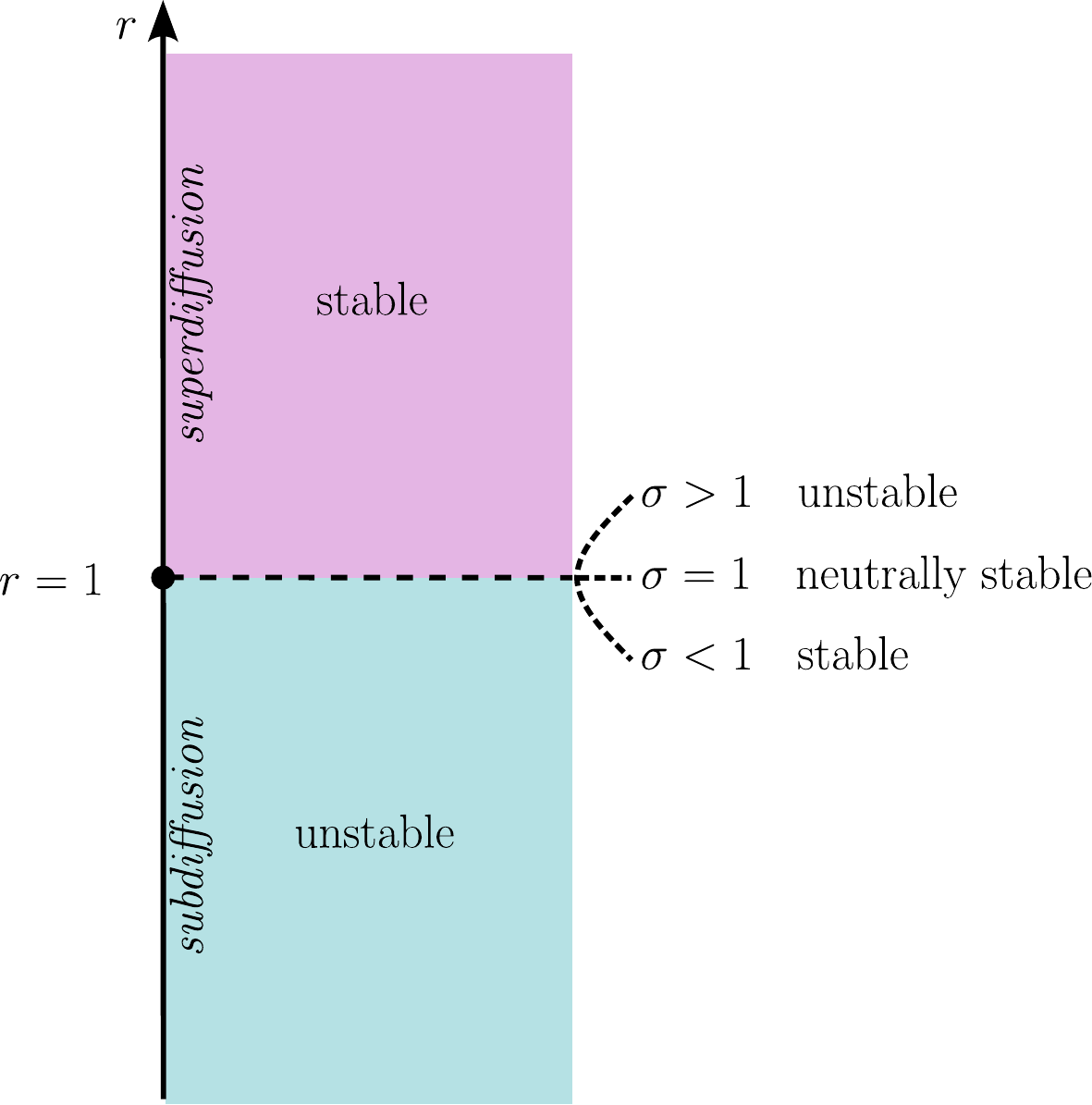}
\caption{\label{fig1}Scheme of the linear instability conditions for different $r$ and $\sigma$.}
\end{figure}

There is a simple solution that can be achieved in the limiting case of creeping flow when viscous effects are dominant over diffusion. Such a solution is obtained by taking the limit $\Sc \to \infty$ in \eqref{15}. In this limiting case, the solution of \eqref{15} is expressed as
\eqn{
f(t) = a_{n,k}^2 \, \sigma \, h(t) \qc h(t) = h(0) \, e^{a_{n,k}^2 \left( \sigma\, t - t^r \right)} ,
}{17}
where $\sigma$ is a neutral stability parameter defined as
\eqn{
\sigma = \frac{k^2 \Ra}{a_{n,k}^6} .
}{18}
The behaviour of $f(t)$ and $h(t)$ at large times is firstly determined by $r$ and, if and only if $r=1$, it is influenced by the neutral stability parameter $\sigma$. In fact, equation \eqref{17} yields
\eqn{
\lim_{t \to +\infty} 
\mqty(
f(t) \\  h(t)
) =
\begin{cases}
\infty \qc&\qfor r<1 ,\\
0 \qc&\qfor r>1 .
\end{cases}
}{19}
For the case of standard diffusion, $r=1$, both $f(t)$ and $h(t)$ for $t \to +\infty$ tend to $0$ if $\sigma<1$, they are stationary for $\sigma=1$, while they tend to infinity if $\sigma>1$. This is the well-known behaviour for the Rayleigh-B\'enard instability reported in many books such as Drazin and Reid \cite{drazin2004hydrodynamic}, Straughan \cite{straughan2013energy} and Barletta \cite{barletta2019routes}. Here, the distinctive feature is the extreme sensitivity of the large time behaviour of disturbances to even minimal departures from $r=1$. Every uncertainty in the determination of $r$ around $1$ may turn stability into instability or vice versa. Such a cornerstone feature has been stressed in similar terms for the case of convective instability in a fluid saturated porous medium by Barletta \cite{barletta2023rayleigh}. A schematic outline of the stability/instability with different $r$ and $\sigma$ is reported in Fig.~\ref{fig1}. 

It will become clear from the analysis carried out in the forthcoming section~\ref{finschnum} that the scheme displayed in Fig.~\ref{fig1} applies also to a finite Schmidt number.

\subsection{A finite Schmidt number}\label{finschnum}
The system of first-order differential equations \eqref{15} can be rewritten as a single second-order differential equation in the unknown $h(t)$,
\eqn{
\dv[2]{h}{t} + a_{n,k}^2 \qty(\Sc + r\, t^{r-1}) \dv{h}{t} + a_{n,k}^2 \big[a_{n,k}^2 r\, \Sc\, t^{r-1}
\nonumber\\
\hspace{5.7cm}+\, r (r-1) t^{r-2} - a_{n,k}^2 \sigma\, \Sc \big] h = 0 .
}{20}
There is just one case where the differential equation \eqref{20} has constant coefficients. It is the case of standard diffusion, $r=1$.

We mention that the test for the instability is based on the large-time behaviour of the solutions of \eqref{20}. Such an ordinary differential equation is homogeneous, meaning that one may have infinite solutions differing just by an overall constant scale factor. In the following, the challenge we pursue is determining the conditions leading to the existence of solutions of \eqref{20} that undergo an unbounded amplification in time when $t \to +\infty$. In fact, this behaviour means instability, while linear stability is found in a parametric range where all possible solutions of \eqref{20} are damped in time, with $h(t)$ tending to zero  when $t \to +\infty$.

\subsubsection{Subdiffusion $(r < 1)$}
With subdiffusion, the large-time approximation of \eqref{20} is a constant coefficient equation given by
\eqn{
\dv[2]{h}{t} + a_{n,k}^2 \Sc \, \dv{h}{t} - a_{n,k}^4 \sigma\, \Sc \, h = 0 .
}{21}
The general solution of \eqref{21} is
\eqn{
h(t) = \eta_1 \exp\qty[-\frac{a^2 \Sc\,t}{2} \left(\sqrt{\frac{4 \sigma}{\Sc} + 1} + 1\right)] + \eta_2 \exp\qty[\frac{a^2 \Sc\,t}{2} \left(\sqrt{\frac{4 \sigma}{\Sc} + 1} - 1\right)],
}{22}
where $\eta_1$ and $\eta_2$ are integration constants.
If $\sigma>0$, there are exponentially growing modes at large $t$. If $\sigma<0$, one has $\Ra < 0$ and, hence, a stably stratified basic state. In this case, \eqref{22} predicts decaying modes at large times either monotonically, if $- \Sc/4 < \sigma < 0$, or oscillatorily, if $\sigma < - \Sc/4$. Thus, our conclusion is that subdiffusion predicts exponentially growing normal modes at large times whenever the basic flow is unstably stratified ($\sigma>0$ or, equivalently, $\Ra>0$). 

\subsubsection{Standard diffusion $(r = 1)$}
For the case of standard diffusion, \eqref{20} simplifies to
\eqn{
\dv[2]{h}{t} + a^2 (\Sc + 1) \, \dv{h}{t} - a^4 \Sc\, \qty(\sigma - 1)\, h = 0,
}{23}
so that the general solution is given by
\eqn{
h(t) = \eta_3 \exp\qty[-\frac{a^2 t}{2} \left(\Sc + 1 + \sqrt{\Sc\, (4 \sigma + \Sc - 2) + 1}\right)]
\nonumber\\
\hspace{4cm}
+\,  \eta_4 \exp\qty[-\frac{a^2 t}{2} \left(\Sc + 1 - \sqrt{\Sc\, (4 \sigma + \Sc - 2) + 1}\right)],
}{24}
where $\eta_3$ and $\eta_4$ are integration constants. From \eqref{24}, one can conclude that there exist exponentially growing modes whenever $\sigma > 1$. Only monotonically decaying modes are allowed if
\eqn{
- \frac{\qty(\Sc - 1)^2}{4 \Sc} < \sigma < 1 .
}{25}
Finally, with a stably stratified basic state ($\sigma<0$ or, equivalently, $\Ra<0$), oscillatorily decaying modes are allowed whenever
\eqn{
\sigma < - \frac{\qty(\Sc - 1)^2}{4 \Sc}  .
}{26}

\subsubsection{Superdiffusion $(r > 1)$}
With superdiffusion, the large-time approximation of \eqref{20} is given by
\eqn{
\dv[2]{h}{t} + a_{n,k}^2 r\, t^{r-1} \, \dv{h}{t} + a_{n,k}^4 r\, \Sc\, t^{r-1} h = 0 .
}{27}
Equation \eqref{27} is independent of the neutral stability parameter $\sigma$ and admits only time-decaying solutions at large times. For example, a possible behaviour is $h(t) \sim e^{-a_{n,k}^2 \Sc\, t}$ which can be gathered from \eqref{27} by neglecting, at large times, $\dd^2 h/\dd t^2$ with respect to either $t^{r-1}\dd h/\dd t$ or $t^{r-1} h$.

\begin{figure}[t]
\centering
\includegraphics[width=0.85\textwidth]{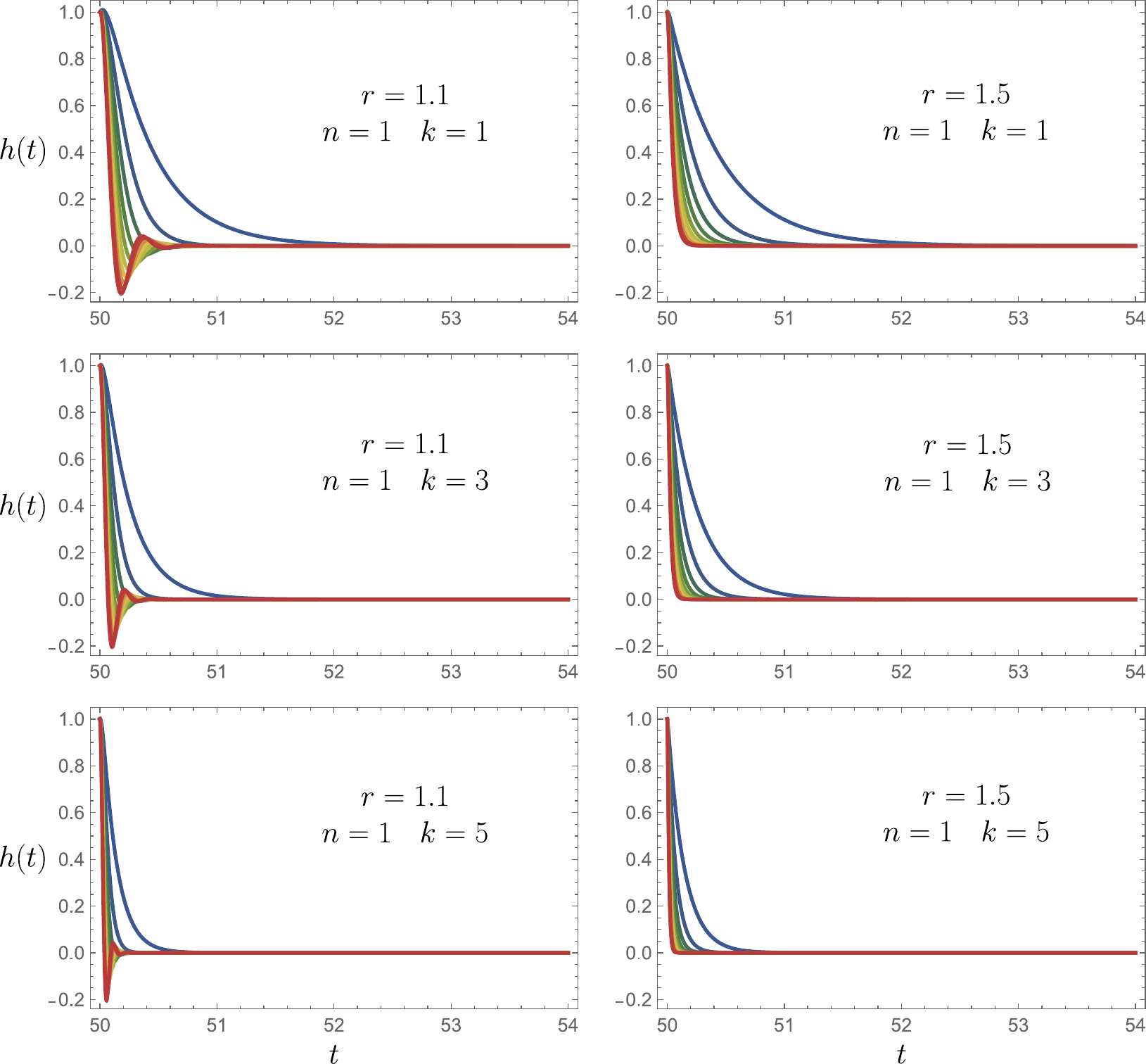}
\caption{\label{fig2}Superdiffusion: plots of the solutions $h(t)$ of \eqref{27} with $t>50$ for different values of $\Sc$ ranging from $0.2$ (blue) to $2$ (red) in steps of $0.2$.}
\end{figure}

Some examples of the behaviour of the solutions $h(t)$ of \eqref{27} are illustrated in Fig.~\ref{fig2}, where the two cases $r=1.1$ and $r=1.5$ are considered with different Schmidt numbers in the range $[0.2, 2]$. Also different modes $(n,k)$ are considered in the different frames. The solutions have been obtained by assuming $t > t_0 = 50$, with $h=1$ and $\dd h/\dd t=1$ at $t=t_0$, where the former condition just fixes the gauge for the solutions of the homogeneous differential equation \eqref{27}. In all cases considered, the fast decay in time of $h(t)$ is quite evident. Such a decay can be either monotonic or oscillatory depending, mainly, on the values of $r$ and $\Sc$. We report that the value assigned to $\dd h/\dd t$ at $t=t_0$ affects the evolution of $h(t)$ at the earlier stages, while the final decay to zero at later times remains unaltered.
We mention that starting the numerical solution at $t = t_0 = 50$ is just a way to address practically the condition of large-times underpinning \eqref{27}.

It is quite important to stress the independence of the time evolution of $h(t)$, at large times, from the neutral stability parameter $\sigma$ and, hence, from the Rayleigh number $\Ra$. As already pointed out, this is testified by \eqref{27}. The physical interpretation is that the large time behaviour of the linear disturbances is unaffected by the buoyancy force, so that the ultimate evolution of the normal modes is diffusion dominated. In other words, the extremely efficient diffusion mechanism of superdiffusion is the key reason of the normal mode damping at large times. Thus, superdiffusion leads to linear stability whatever is the value of $\sigma$ or $Ra$.

\section{Superdiffusion, nonlinear theory, fixed surfaces}
\label{S:SuperNonlin}

We now analyse the fully nonlinear superdiffusion case. The equations are again \eqref{8}, but we rescale
$C$
as
$\phi=RC$,
where
$\Ra=R^2$.
Thus, the nonlinear perturbation equations are
\begin{equation}\label{E:PertEqSuper}
\begin{aligned}
&\frac{\partial U_j}{\partial x_j}=0,\\
&\frac{1}{Sc}\Bigl(\frac{\partial U_i}{\partial t}+U_j\frac{\partial U_i}{\partial x_j}\Bigr)=\nabla^2 U_i+R\phi\delta_{i3}-\frac{\partial P}{\partial x_i}\,,\\
&\frac{\partial \phi}{\partial t}+U_i\frac{\partial \phi}{\partial x_i}=RW+rt^{r-1}\nabla^2\phi.
\end{aligned}
\end{equation}
We now consider the case of two fixed surfaces and so the boundary conditions are
\begin{equation}\label{E:BCsNonlin}
U_i=0,\quad\phi=0,\qquad{\rm on}\,\,z=0,1.
\end{equation}
In addition to \eqref{E:BCsNonlin} the solution is subject to the requirement that
$(U_i,\phi,P)$
satisfy a plane tiling periodicity; such a condition is explained in depth in Chandrasekhar \cite[~pages 43--52]{Chandrasekhar:1981}. The initial conditions are
\begin{equation}\label{E:ICsNonlin}
U_i({\bf x},0)=G_i({\bf x}),\quad\phi({\bf x},0)=\Phi({\bf x}).
\end{equation}

Let
$V$
denote a period cell for a solution to \eqref{E:PertEqSuper}-\eqref{E:ICsNonlin}, and let 
$\Vert\cdot\Vert$
and
$(\cdot,\cdot)$
denote the norm and inner product on
$L^2(V)$.
We wish to analyse the asymptotic behaviour of a solution to \eqref{E:PertEqSuper}-\eqref{E:ICsNonlin} as time
$t\to\infty$.

We first require a general bound for
$\Vert{\bf U}(t)\Vert$
and
$\Vert\phi(t)\Vert$.
Thus, suppose
$r>1$,
then multiply the second equation \eqref{E:PertEqSuper} by
$U_i$
and integrate over 
$V$,
and likewise
multiply the third equation \eqref{E:PertEqSuper} by
$\phi$
and integrate over 
$V$.
Upon employing the boundary conditions one may show
\begin{align}
&\frac{1}{2}\dv{}{t}\Bigl(\frac{1}{\Sc}\Vert{\bf U}\Vert^2+\Vert\phi\Vert^2\Bigr) = 2R(\phi,W)- {\Vert\grad{\bf U}\Vert}^2-rt^{r-1}{\Vert\grad\phi\Vert}^2,\label{E:EnId}\\
&\frac{1}{2}\dv{}{t}\Bigl(\frac{1}{\Sc}\Vert{\bf U}\Vert^2+\Vert\phi\Vert^2\Bigr) \le 2R(\phi,W)- {\Vert\grad{\bf U}\Vert}^2.\label{E:EnIneq}
\end{align}
Next, use the arithmetic-geometric mean inequality on the 
$(\phi,W)$
term in \eqref{E:EnIneq} and Poincar\'e's inequality ${\Vert\grad{\bf U}\Vert}^2 \ge \pi^2 {\Vert{\bf U}\Vert}^2$ to derive from that inequality 
\begin{align}
\dv{}{t}\Bigl(\frac{1}{\Sc}\Vert{\bf U}\Vert^2+\Vert\phi\Vert^2\Bigr) \le&\frac{2R^2}{\pi^2}\Vert\phi\Vert^2 \le \frac{2R^2}{\pi^2}\Bigl(\frac{1}{\Sc}\Vert{\bf U}\Vert^2+\Vert\phi\Vert^2\Bigr).\label{E:IneqPart}
\end{align}
One integrates \eqref{E:IneqPart} to show that for all
$t>0$,
\begin{equation}\label{E:SolnBd}
\frac{1}{\Sc}\Vert{\bf U}(t)\Vert^2+\Vert\phi(t)\Vert^2\le\Bigl(\frac{\Vert{\bf G}\Vert^2}{\Sc}+\Vert\Phi\Vert^2\Bigr)\,\exp\Bigl(\frac{2R^2t}{\pi^2}\Bigr)\,.
\end{equation}

To derive an asymptotic solution estimate we return to equation \eqref{E:EnId} and then use Poincar\'e's inequality
${\Vert\grad\phi\Vert}^2\ge\pi^2\Vert\phi\Vert^2,$
and the arithmetic-geometric mean inequality to derive the following inequality:
\begin{equation}\label{E:MidInt}
\frac{1}{2}\dv{}{t}\Bigl(\frac{1}{\Sc}\Vert{\bf U}\Vert^2 + \Vert\phi\Vert^2\Bigr)+\frac{\pi^2}{2}\Vert{\bf U}\Vert^2
+\Bigl(rt^{r-1}\pi^2-\frac{2R^2}{\pi^2}\Bigr)\Vert\phi\Vert^2\le 0,
\end{equation}
for all 
$t>0$.

We now require
$t>t_0=(2R^2/r\pi^4)^{1/(r-1)},$
and then integrate inequality \eqref{E:MidInt} with respect to time from
$2t_0$ to $T>2t_0$.
In this way we find
\begin{align}
\frac{\pi^2}{2}\int_{2t_0}^{T}\Vert{\bf U}(t)\Vert^2\,\dd t &+ \int_{2t_0}^{T}\Bigl(r\pi^2t^{r-1}-\frac{2R^2}{\pi^2}\Bigr)\Vert\phi(t)\Vert^2 \dd t\notag\\
&\le \frac{1}{2\Sc}\Vert{\bf U}(2t_0)\Vert^2+\frac{1}{2}\Vert\phi(2t_0)\Vert^2 - \frac{1}{2\Sc}\Vert{\bf U}(T)\Vert^2-\frac{1}{2}\Vert\phi(T)\Vert^2\notag\\
&\le \frac{1}{2\Sc}\Vert{\bf U}(2t_0)\Vert^2+\frac{1}{2}\Vert\phi(2t_0)\Vert^2 \notag\\
&\le\frac{1}{2\Sc}\Bigl(\Vert{\bf G}\Vert^2+\Vert\Phi\Vert^2\Bigr)\,\exp\,\Bigl[\frac{4R^2t_0}{\pi^2}\Bigr]\,\equiv\,K<\infty\,.\label{E:UpBd}
\end{align}
The constant
$K$
defined by the expression in \eqref{E:UpBd} is independent of $T$, and inequality \eqref{E:UpBd} holds for all
$T>2t_0$. The coefficient of the fourth term on the left of \eqref{E:UpBd} satisfies
\begin{equation*}
r\pi^2t^{r-1}-\frac{2R^2}{\pi^2}\ge
r\pi^2(2t_0)^{r-1}-\frac{2R^2}{\pi^2}\equiv \chi >0.
\end{equation*}
Hence, from \eqref{E:UpBd}, one can let $T \to \infty$ to obtain
\begin{equation}\label{E:NmUpBd}
\int_{2t_0}^{\infty}\Vert{\bf U}(t)\Vert^2 \dd t<\frac{2K}{\pi^2}\,,\quad {\rm and}\quad
\int_{2t_0}^{\infty}\Vert\phi(t)\Vert^2 \dd t<\frac{K}{\chi}\,,
\end{equation}
so that both integrals are bounded.

Now, we observe from inequality \eqref{E:MidInt} that for
$t\ge 2t_0$,
\begin{equation}\label{E:DerivBdd}
\dv{}{t}\Bigl(\frac{1}{\Sc}\Vert{\bf U}\Vert^2+\Vert\phi\Vert^2\Bigr)\le 0.
\end{equation}
Hence, from \eqref{E:NmUpBd} and \eqref{E:DerivBdd} we deduce
$\Vert{\bf U}(t)\Vert^2\to 0,$ 
$\Vert\phi(t)\Vert^2\to 0,$ 
as
$t\to\infty$,
regardless of the size of the Rayleigh number 
$R^2$.

\section{Subdiffusion, fixed surfaces}
\label{S:SubFixed}

The perturbation equations for subdiffusion are the same as \eqref{E:PertEqSuper}, together with the boundary and initial conditions \eqref{E:BCsNonlin},
\eqref{E:ICsNonlin}. Now, the anomalous diffusion term in \eqref{E:PertEqSuper},
$r t^{r-1}\nabla^2\phi$, features $0<r<1$.

In this case we deal with the linearized system of equations arising from \eqref{E:PertEqSuper}, namely
\begin{equation}\label{E:PertSub}
\begin{aligned}
&\frac{\partial U_i}{\partial x_i}=0,\\
&\frac{1}{\Sc}\frac{\partial U_i}{\partial t}=-\frac{\partial P}{\partial x_i}+R\phi\delta_{i3}+\nabla^2 U_i,\\
&\frac{\partial \phi}{\partial t}=RW+r t^{r-1}\nabla^2\phi.
\end{aligned}
\end{equation}
To proceed with an analysis of asymptotic behaviour for a solution to \eqref{E:PertSub}, \eqref{E:BCsNonlin}, \eqref{E:ICsNonlin},
we define the function
$F(t)$ by
\begin{equation}\label{E:FDef}
F(t)=\frac{1}{\Sc}\Vert{\bf U}(t)\Vert^2+\Vert\phi(t)\Vert^2.
\end{equation}
Differentiate
$F$
and use \eqref{E:PertSub}, integration by parts and the boundary conditions to see that
\begin{align}
\dv{F}{t}&=\frac{2}{\Sc}\qty(U_i, \pdv{U_i}{t})+2\qty(\phi, \pdv{\phi}{t}),\label{E:Fprime}\\
&=2\qty(U_i,-\pdv{P}{x_i}+R\phi\delta_{i3}+\nabla^2 U_i)+2\qty(\phi,RW+r t^{r-1}\nabla^2\phi),\notag\\
&=4R(\phi,W)-2{\Vert\grad{\bf U}\Vert}^2-2r t^{r-1}{\Vert\grad\phi\Vert}^2.\label{E:Fprime2}
\end{align}
Perform a further differentiation to deduce
\begin{align}
\dv[2]{F}{t}=&\ 4R\qty(\pdv{\phi}{t},W)+4R\qty(\phi, \pdv{W}{t})-4\qty(\pdv{U_i}{x_j}, \pdv[2]{U_i}{x_j}{t}) - 4r t^{r-1}\qty(\pdv{\phi}{x_i},\pdv[2]{\phi}{x_i}{t})\notag\\
-&\ 2r(r-1)t^{r-2}{\Vert\grad\phi\Vert}^2\notag\\
=&\ 4\qty(\pdv{\phi}{t},RW + r t^{r-1}\nabla^2\phi)+4\qty(\pdv{U_i}{t}, R\phi\delta_{i3} + \nabla^2 U_i)-2r(r-1)t^{r-2}{\Vert\grad\phi\Vert}^2 \notag\\
=&\ 4\left\Vert\pdv{\phi}{t}\right\Vert^2+\frac{4}{\Sc}\left\Vert\pdv{\bf U}{t}\right\Vert^2+4\qty(\pdv{U_i}{t},\pdv{P}{x_i})-2r(r-1)t^{r-2}{\Vert\grad\phi\Vert}^2 \notag\\
=&\ 4\left\Vert\pdv{\phi}{t}\right\Vert^2+\frac{4}{\Sc}\left\Vert\pdv{\bf U}{t}\right\Vert^2 +2r(1-r)t^{r-2}{\Vert\grad\phi\Vert}^2 \ge 4\left\Vert\pdv{\phi}{t}\right\Vert^2+\frac{4}{\Sc}\left\Vert\pdv{\bf U}{t}\right\Vert^2.\label{E:F2prime}
\end{align}
Define
$S^2$
by
\begin{equation}
S^2=
\qty(\frac{1}{\Sc}\Vert{\bf U}\Vert^2+\Vert\phi\Vert^2)
\qty(\frac{1}{\Sc}\left\Vert\pdv{\bf U}{t}\right\Vert^2+\left\Vert\pdv{\phi}{t}\right\Vert^2)
-\qty[\frac{1}{\Sc}\qty(U_i, \pdv{U_i}{t})+ \qty(\phi, \pdv{\phi}{t})]^2\,,
\end{equation}
and note that
$S^2\ge 0$
by the Cauchy-Schwarz inequality. 
Then, employ \eqref{E:FDef}, \eqref{E:Fprime} and \eqref{E:F2prime} to see that
\begin{equation}\label{E:Conv}
F \dv[2]{F}{t}-\qty(\dv{F}{t})^2\ge 0,
\end{equation}
for all
$t>0$.
We assume the initial data are non-zero so that 
$F^2>0$
in some interval beyond
$t=0$. 
Then, divide \eqref{E:Conv} by
$F^2$
to see that
\begin{equation}\label{E:LogConv}
\dv[2]{(\log F)}{t}\ge 0.
\end{equation}
This inequality is integrated twice and then the exponential is taken of the result to find
\begin{equation}\label{E:Fest}
F(t)\ge F(0)\,\exp\!\qty[\frac{t}{F(0)}\left.\dv{F}{t}\right|_{t=0}]\,.
\end{equation}
The form of the right hand side of \eqref{E:Fest} ensures
$F>0$
for all
$t$.
Furthermore, using the initial data and the differential equations
\begin{equation}\label{E:FPIn}
\left.\dv{F}{t}\right|_{t=0}=-2{\Vert\grad{\bf G}\Vert}^2+4R(\Phi,G_3).
\end{equation}
No matter what value 
$R$
has, there are initial data such that
$\left.\dd F/\dd t\right|_{t=0}>0$.
Thus, \eqref{E:Fest} shows that a solution to \eqref{E:PertSub}, \eqref{E:BCsNonlin} and \eqref{E:ICsNonlin} in the sub-diffusion case will grow
exponentially for
$t$
large enough.

We have thus demonstrated by rigorous analysis that a solution to the diffusion problem for the Navier-Stokes equations will behave in a similar manner
to that found in Section~\ref{antnormod} with a normal mode analysis, as well as in Barletta \cite{barletta2023rayleigh} for the analogous problem in a Darcy porous material with no inertia. In the super diffusion
case for the Navier-Stokes equations the solution may grow relatively large for a finite time, but will eventually decay to zero. For the sub-diffusion
case the solution may also increase or decay for a finite time, but then for the linearized problem the solution will grow at least exponentially as
$t\to\infty$.

\vskip12pt

\subsubsection*{Remark} 
One may derive analogous results to those in sections \ref{S:SuperNonlin} and \ref{S:SubFixed} 
for a solution to the Navier-Stokes-Voigt equations
with anomalous diffusion of the solute field. In this case, the perturbation equations have form
\begin{align*}
&\pdv{U_i}{x_i}=0,\\
&\frac{1}{\Sc}\qty(\pdv{U_i}{t}+U_j\, \pdv{U_i}{x_j}) - \lambda\,\nabla^2 \pdv{U_i}{t} = - \pdv{P}{x_i} + R\, \phi\, \delta_{i3}+\nabla^2 U_i,\\
&\pdv{\phi}{t}+U_i\, \pdv{\phi}{x_i} = R\, W + r\,t^{r-1}\, \nabla^2\phi.
\end{align*}
For the superdiffusion case, one shows that 
$\Vert{\bf U}\Vert,$ ${\Vert\grad{\bf U}\Vert}$
and
$\Vert\phi\Vert$
must all decay to zero as
$t\to\infty$.
For the subdiffusion case, one derives an exponential estimate in the measure
$\Sc^{-1}\Vert{\bf U}\Vert^2+\lambda{\Vert\grad{\bf U}\Vert}^2+\Vert\phi\Vert^2.$

\section{Conclusions}
The Rayleigh-B\'enard instability in a horizontal binary-mixture fluid layer has been revisited by taking into account the effects devised in the anomalous diffusion model. Such effects are caused by departures from the linear time evolution of the molecular position variance predicted by the standard mass diffusion theory. A power-law time dependence holds instead, leading to either subdiffusion or superdiffusion regimes that occur when the time evolution is sublinear or superlinear, respectively. The mass diffusion equation features a time-dependent diffusion coefficient so that the standard techniques employed for the linear or nonlinear analysis of the Rayleigh-B\'enard instability have to be entirely reconsidered. This paper offered one of the first theoretical schemes of how such a stability/instability analysis should be approached. In fact, there exists just a couple of previous investigations sharing the same objective \cite{barletta2023rayleigh, barstra2023rayleigh}. However, they are relative to the case of Darcy's flow in a porous material. 

The modal analysis of the linearised equations for the disturbances has been carried out leading to the formulation of an initial value problem in time to be solved in order to establish the stability/instability of the equilibrium state depending on the subdiffusive, standard or superdiffusive nature of the mass transfer. This analysis has shown that the existence of a neutral stability condition and a critical value of the Rayleigh number $\Ra$ is just typical of standard mass diffusion whereas subdiffusion means instability and superdiffusion means stability for every $\Ra>0$. 

The energy method study of the superdiffusion and subdiffusion cases confirmed the results obtained by the modal analysis. In particular, the energy method study of superdiffusion provided an extension of the results obtained through the modal analysis to the nonlinear case and to the rigid wall version of the velocity boundary conditions.

An evident feature of the anomalous diffusion theory of instability is that it generally rules out the formulation of an eigenvalue problem as the basis for the determination of the instability parametric threshold. On the other hand, the stable/unstable behaviour of the equilibrium state is ascertained by inspecting the large time evolution of a differential equation in time. How this fundamental difference leads to a new systematic scheme for the stability analysis is yet to be investigated and offers a challenge for future research.\vskip6pt

\vskip24pt

\subsubsection*{Acknowledgements}
{The work by Antonio Barletta was supported by Alma Mater Studiorum Universit\`a di Bologna, grant number RFO-2023.
The work of Brian Straughan was supported by the Leverhulme Trust, grant number EM/2019-022/9.}


\end{document}